\documentclass[5p,times,square,sort,compress,comma,10pt]{elsarticle}
\usepackage{multirow,setspace,times,amssymb,amsmath,graphicx,color,rotating,subfigure,url}
\usepackage{lineno}
\usepackage{dcolumn,multirow}
\usepackage[]{natbib}
\journal{Chaos, Solitons \& Fractals}

\begin{document}

\begin{frontmatter}

\title{Statistical properties of user activity fluctuations in virtual worlds}
\author[SB,RCE]{Yan-Hong Yang}
\author[SB,RCE]{Wen-Jie Xie}
\author[RCE,SSSE]{Ming-Xia Li}
\author[SB,RCE]{Zhi-Qiang Jiang\corref{cor}}
\ead{zqjang@ecust.edu.cn} %
\author[SB,SSSE,SS]{Wei-Xing Zhou\corref{cor}}
\cortext[cor]{Corresponding authors. Address: 130 Meilong Road, P.O. Box 114, School of Business, East China University of Science and Technology, Shanghai 200237, China, Phone: +86 21 64253634, Fax: +86 21 64253152.}
\ead{wxzhou@ecust.edu.cn} %

\address[SB]{School of Business, East China University of Science and Technology, Shanghai 200237, China}
\address[RCE]{Research Center for Econophysics, East China University of Science and Technology, Shanghai 200237, China}
\address[SSSE]{School of Sports Science and Engineering, East China University of Science and Technology, Shanghai 200237, China}
\address[SS]{Department of Mathematics, East China University of Science and Technology, Shanghai 200237, China}

\begin{abstract}
  User activity fluctuations reflect the performance of online society. We investigate the statistical properties of 1-min user activity time series of simultaneously online users inhabited in 95 independent virtual worlds. The number of online users exhibits clear intraday and weekly patterns due to human's circadian rhythms and week cycles. Statistical analysis shows that the distribution of absolute activity fluctuations has a power-law tail for 44 virtual worlds with an average tail exponent close to 2.15. The partition function approach unveils that the absolute activity fluctuations possess multifractal features for all the 95 virtual worlds. For the sample of 44 virtual worlds with power-law tailed distributions of the absolute activity fluctuations, the width of singularity $\Delta\alpha$ is negatively correlated with the maximum activity ($p$-value=0.070) and the time to the maximum activity ($p$-value=0.010). The negative correlations are not observed for neither the other 51 virtual worlds nor the whole sample of the 95 virtual worlds. In addition, numerical experiments indicate that both temporal structure and large fluctuations have influence on the multifractal spectrum. We also find that the temporal structure has stronger impact on the singularity width than large fluctuations.
  
[{\textit{Chaos, Solitons \& Fractals}} {\bf{105}}, 271-278 (2017)] 

%
\end{abstract}


\end{frontmatter}

\section{Introduction}
\label{intro}

A massive multiplayer online role-playing game (MMORPG) forms an online virtual world, where people can work and interact with one another in a somewhat realistic manner. Therefore, virtual worlds have great potential for research in the social, behavioral, and economic sciences \cite{Bainbridge-2007-Science}. For instance, we can embed evolutionary games in a virtual world to study the formation of human cooperation \cite{Grabowski-Kosinski-2008-APPA} and to understand the evolution of wealth distribution \cite{Tseng-Li-Wang-2010-EPJB}. A pioneering work was done by Castronova, who traveled in a virtual world called ``Norrath'' and performed preliminary analysis of its economy \cite{Castronova-2001-WP}. Recently, there have been also efforts in the field of computational social sciences from a complex network perspective \cite{Grabowski-2007-PA,Grabowski-2009-PA,Grabowski-Kosinski-2008-APPB,Grabowski-Kruszewska-2007-IJMPC,Grabowski-Kruszewska-Kosinski-2008-EPJB,Grabowski-Kruszewska-Kosinski-2008-PRE,Xie-Li-Jiang-Tan-Podobnik-Zhou-Stanley-2016-SR}. In addition to its scientific potentials, virtual worlds could act as nice places for real social activities, such as marketing \cite{Matsuda-2003-Presence,Castronova-2005-HBR,Hemp-2006-HBR}, and provide opportunities for players to make real money \cite{Papagiannidis-Bourlakis-Li-2008-TFSC}.

The number of instant online users is an important indicator for scientific and commercial purposes. The number of registered users is closely related to the profit of an MMORPG company and the instant number of online users shows the degree of popularity of an MMORPG \cite{Jiang-Ren-Gu-Tan-Zhou-2010-PA}. The number of instant online users is an analogue to various instant society flows \cite{Bogachev-Bunde-2009-EPL,Cai-Fu-Zhou-Gu-Zhou-2009-EPL}. Moreover, we note that the online-offline activities of users have the power to identify game cheaters and the gaming session durations of the majority of normal users are distributed according to the Weibull distribution \cite{Jiang-Zhou-Tan-2009-EPL}, which deviates the power-law bursts of human activities in many social systems \cite{Barabasi-2005-Nature}. In addition, power-law behavior extensively exist in social and natural sciences \cite{Jiang-Xie-Li-Podobnik-Zhou-Stanley-2013-PNAS,Clauset-Shalizi-Newman-2009-SIAMR}, which is identified in our investigated data. In a word, it is meaningful to study the linear and nonlinear dynamics of the number of instant online users and duration between login and logoff moments. We mainly focus on the multifractal nature of the absolute fluctuations of user activities (the absolute increments of instant numbers of simultaneously online users) in this work.

Multifractals is ubiquitous in natural and social sciences \cite{Mandelbrot-1983}. Many different methods have been applied to characterize the hidden multifractal behavior of different social variables, such as the fluctuation scaling analysis \cite{Eisler-Kertesz-2007-EPL,Jiang-Guo-Zhou-2007-EPJB}, the structure function method \cite{Kolmogorov-1962-JFM,VanAtta-Chen-1970-JFM,Anselmet-Gagne-Hopfinger-Antonia-1984-JFM,Ghashghaie-Breymann-Peinke-Talkner-Dodge-1996-Nature}, the multifractal detrended fluctuation analysis (MF-DFA) \cite{CastroESilva-Moreira-1997-PA,Weber-Talkner-2001-JGR,Kantelhardt-Zschiegner-KoscielnyBunde-Havlin-Bunde-Stanley-2002-PA}, the multifractal detrending moving average analysis (MF-DMA) \cite{Gu-Zhou-2010-PRE}, the partition function method \cite{Grassberger-1983-PLA,Hentschel-Procaccia-1983-PD,Grassberger-1985-PLA,Halsey-Jensen-Kadanoff-Procaccia-Shraiman-1986-PRA,Xie-Jiang-Gu-Xiong-Zhou-2015-NJP}, the multiplier method \cite{Chhabra-Sreenivasan-1992-PRL,Jouault-Lipa-Greiner-1999-PRE,Jiang-Zhou-2007-PA}, the wavelet transform approaches \cite{Muzy-Bacry-Arneodo-1991-PRL,Muzy-Bacry-Arneodo-1993-PRE}, and the microcanonical multifractal analysis \cite{Turiel-Yahia-PerezVicente-2008-JPA,Pont-Turiel-PerezVicente-2009-PA}, some of which are borrowed from the multifractal analysis of turbulence data. We apply the partition function approach to the absolute fluctuation time series of 1-min online user number to uncover the multifractal nature of the records in the present study.

The rest of this paper is organized as follows. Section \ref{S1:Data} describes the data used in our study, including the time series of user activities and its fluctuations. Section \ref{S1:Statistics} investigates the intraday patterns and weekly patters of user activities and Section \ref{S1:PDF} studies the probability distribution of the fluctuations of user activities $|\Delta N|$. We perform multifractal analysis of the user activity fluctuations based on the partition function approach and unfold the relationships between multifractal nature and the performance of virtual worlds. We summarize our findings in Section \ref{S1:Conclusion}.

\section{Data description}
\label{S1:Data}

We use a huge database recorded from 95 servers of a popular MMORPG in China to uncover the patterns characterizing virtual worlds. Our data set contains all in-game action logs for 111 days from May 16 to September 4 in 2011. However, we mainly focus on the online-offline logs in this study. An entry is written to the log file when a user goes offline. Therefore, the entries in a log file are arranged according to an increasing order of logout moments. Each entry contains three pieces of information: the masked user ID, its login time, and its logout time. The resolution of the time stamps is 1 second. For each user, we collect all the associated entries. During this period, on average, there were more than 100 000 users created on each server. For security sake, the true user IDs have been encrypted into numbers from 1 to the ordinal number of the last ID for each virtual world.

We use 1-min number $N_i(t)$ of simultaneously online users as the user activity of the $i$-th virtual world. Considering the privacy of the data, we define a quantity $n_i(t)$  as a substitute for $N_i(t)$, which does not change the results,
\begin{equation}
  n_i(t)=N_i(t)/N_{\max},
  \label{Eq:ni:t}
\end{equation}
where
\begin{equation}
  N_{\max} = \max\limits_i\{N_{i,\max}, i=1,2, \cdots, 95\}
  \label{Eq:N:max}
\end{equation}
in which $N_{i,\max}$ is the maximum of the user activities $N_i(t)$ of the $i$-th virtual world:
\begin{equation}
  N_{i,\max} = \max\limits_t\{N_i(t), t=1,2, \cdots, T\}
  \label{Eq:Ni:max}
\end{equation}
Accordingly, the relative maximum 1-min number of online users can be calculated as follow:
\begin{equation}
  n_{i,\max}=N_{i,\max}/N_{\max}.
  \label{Eq:ni:max}
\end{equation}
We find that the majority of $n_{i,\max}$ are greater than 0.8 and the mean is 0.8105, which indicates that there exist small differences in the maximum activity $N_{i,\max}$ among most virtual worlds. Meanwhile, we have removed the abnormal activities (e.g. when the servers were scheduled for maintaining or during game version updating) of the 95 virtual worlds in order to ensure statistical significance.

\begin{figure}[!htb]
  \centering
  \includegraphics[width=8cm]{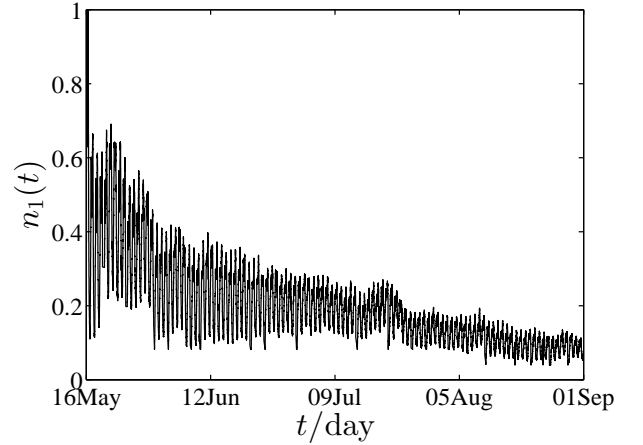}
  \caption{Time series of the relative number $n_1(t)$ of online users minute by minute for a typical virtual world.}
  \label{Fig:MMORPG:n:Evolution}
\end{figure}

Fig.~\ref{Fig:MMORPG:n:Evolution} illustrates the evolution of 1-min online user numbers for a typical server during the period under investigation. The maximum relative activity $n_{i,\max}$ is reached at the beginning of the recording period. Especially, there exist two evident local humps in the plot around 2011/05/21 and 2011/07/20. These humps are mainly caused by some new marketing actions organized by the online game operators. We find that other curves almost share the same shape as in Fig.~\ref{Fig:MMORPG:n:Evolution} except for some special dates, and the rest time series of the virtual worlds also have similar features.

\begin{figure}[!htb]
  \centering
  \includegraphics[width=8cm]{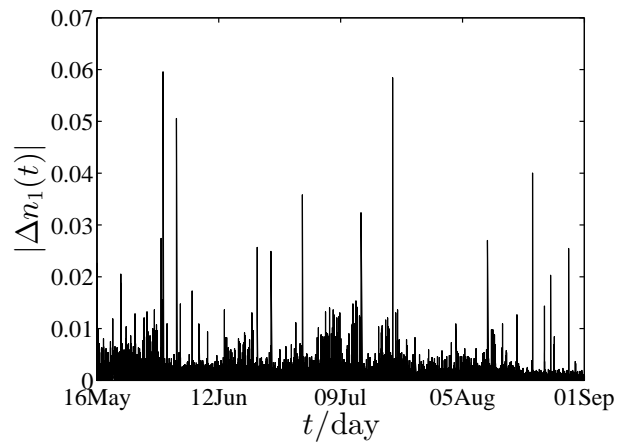}
  \caption{Evolution of the fluctuations $|\Delta{n}_1(t)|$ of user activities in the same typical virtual world used in Fig.~\ref{Fig:MMORPG:n:Evolution}.}
  \label{Fig:MMORPG:dn:Evolution}
\end{figure}

Fig.~\ref{Fig:MMORPG:dn:Evolution} illustrates the evolution of absolute fluctuations of online user activities, which is the absolute difference
\begin{equation}
  \Delta{n}_1(t)=|n_1(t)-n_1(t-1)|.
  \label{Eq:MMORPG:Dn}
\end{equation}
One can find that the time series exhibits large fluctuations and intermittent behavior. In addition, Eq.~(\ref{Eq:MMORPG:Dn}) is a substitute of Eq.~(\ref{Eq:MMORPG:DN}), which does not change the results.
\begin{equation}
  \Delta{N}_1(t)=|N_1(t)-N_1(t-1)|.
  \label{Eq:MMORPG:DN}
\end{equation}

\section{Intraday pattern and weekly pattern}
\label{S1:Statistics}

In order to investigate the seasonal patterns in the time series of the online user activities, we calculate the average number $A_i(\tilde{t})$ of online users as follows
\begin{equation}\label{Eq:Intraday:Patterns}
  A_i(\tilde{t})=\frac{1}{M_i}\sum_{j=1}^{M_i} {n}_i^j(\tilde{t}),
\end{equation}
where $i=1,2,3,\cdots,95$ and $M_i$ is the number of operating days in the $i$-th virtual world, ${n}_{i}^{j}(\tilde{t})$ is the 1-min relative number of online users, which is divided by its maximum at time $\tilde{t}$ of day $j$ as defined in Eq.~(\ref{Eq:ni:t}).

\begin{figure}[!htp]
  \centering
  \includegraphics[width=8cm]{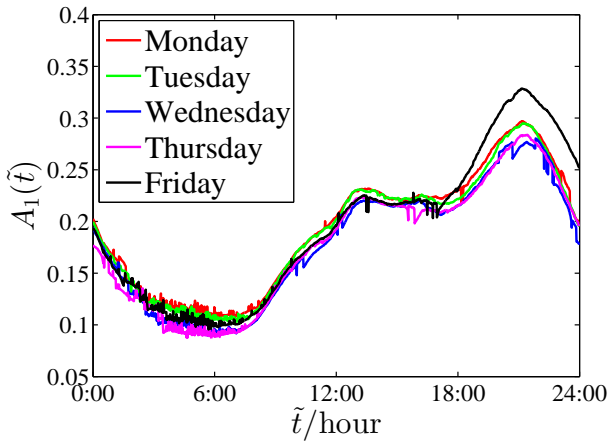}
  \caption{Intraday patterns for Monday, Tuesday, Wednesday, Thursday, and Friday (consider only the sample of working day) of online user numbers. Note that the Friday curve is slight different from other curves. Other virtual worlds have very similar patterns.}
  \label{Fig:MMORPG:SP:Weekday}
\end{figure}

We first determine the intraday patterns on working days respectively for Monday, Tuesday, Wednesday, Thursday and Friday, as presented in Fig.~\ref{Fig:MMORPG:SP:Weekday}. Roughly speaking, the five curves almost overlap and no remarkable differences are observed among these days. On average, the maximum number of online users is reached at around 21:00 p.m. after finishing dinner and before going to bed. After that the number of online users decreases gradually till about 6:00 a.m. on the next day. The time of low activity in virtual worlds is reminiscent of cell phone users in reality \cite{Jiang-Xie-Li-Podobnik-Zhou-Stanley-2013-PNAS,Jiang-Xie-Li-Zhou-Sornette-2016-JSM}. Early in the morning around 5:00 a.m., players start to enter virtual worlds again and the number of online users increases. This increasing trend ends till 21:00 p.m., except for the afternoon during which the number of online users exhibits a plateau. The majority of players are young college students and young workers\cite{Shen-Chen-2015-SN}. Generally speaking, they would like to finish their real-world tasks in the afternoon, getting ready for night or overnight game-playing. In addition, the relatively sharp drop of online user of the weekend curve is probably due to the fact that most of the player have to sleep normally in the Sunday evening so that they can have a normal life on Monday.

A careful scrutiny unveils that the average number on Friday increases quicker after 17:00 and remains above other four curves from 18:00 to 24:00 for all virtual worlds, which is slightly different from the result of another game in 2007 in which the prominence of the Friday curve started after lunch \cite{Jiang-Ren-Gu-Tan-Zhou-2010-PA}. The intraday pattern in Friday evening is explained by the fact that Fridays are followed by Saturdays and most of the players are free in weekends, while that in the Friday afternoon 17:00 is explained by the fact that most college students do not have courses and many official institutions have much less work to do, for instance, only a small part of the officials might have obligations. This Friday afternoon 17:00 pattern is expected to be idiosyncratic for MMORPGs played mainly by Chinese people. We also notice a sharp discontinuity in the Friday curve from the right end to the left end, which is trivial since the next moment after the midnight of Friday is before dawn on Saturday, which has higher activity than working days.

\begin{figure}[!htp]
  \centering
  \includegraphics[width=8cm]{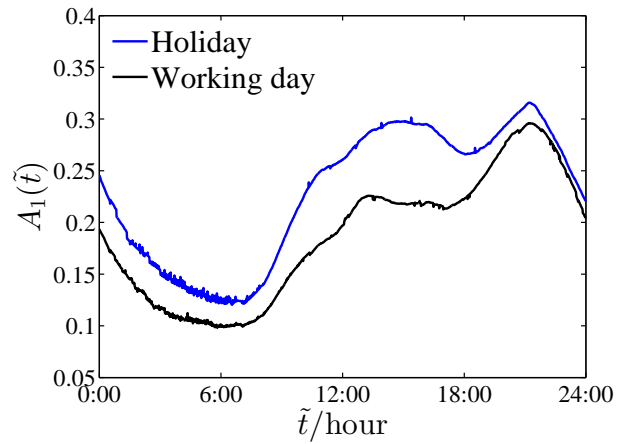}
  \caption{(color online) Intraday patterns of online user numbers on working days and holidays (including weekends and public holidays) for a typical virtual world. It is trivial that there are more users playing in virtual worlds on weekends and public holidays. The most significant difference between the two patterns lies in 17:00-18:00. Other virtual worlds have very similar patterns.}
  \label{Fig:MMORPG:SP:Weekend}
\end{figure}

We then partition all the 111 days into two groups, one containing all working days and the other including all weekends and public holidays. The intraday patterns of these two groups of days are shown in Fig.~\ref{Fig:MMORPG:SP:Weekend}. At a first glance, the intraday patterns in weekdays and in weekends are quite similar, except that the users are more active on the weekends. The trend of online user number is consistent with the circadian rhythm of human activities. However, a significant difference appears from 15:00 p.m. to 18:00 p.m. between working days and holidays. At least for part of the players, game-playing is only part of their lives and they will hang out for other social activities, such as going shopping and arranging dinner with families, friends or colleagues.

\begin{figure}[!htp]
  \centering
  \includegraphics[width=8cm]{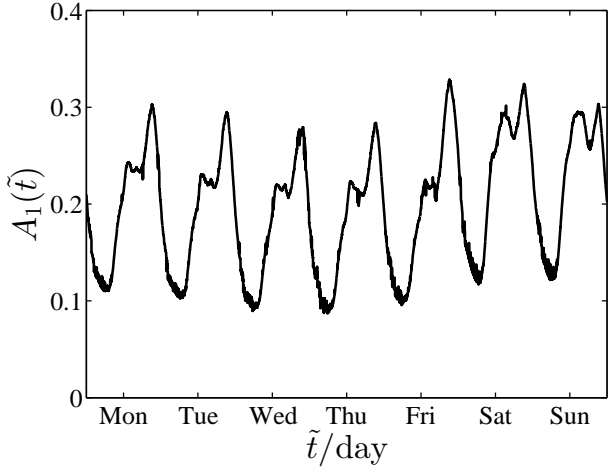}
  \caption{Weekly patterns of online user numbers for a typical virtual world. Although the seven oscillations have very similar shape, there are clear differences between weekdays and weekends, and among weekdays as well. Other virtual worlds have very similar patterns.}
  \label{Fig:MMORPG:SP:Weekly}
\end{figure}

We now turn to investigate the weekly patterns of online user activities. The result is illustrated in Fig.~\ref{Fig:MMORPG:SP:Weekly}. We observe very nice ``periodic'' oscillations on a daily base with mild fluctuations. Each periodic oscillation exhibits two peaks, which is more prominent in the weekend curves. On average, the activity of virtual worlds decreases from Monday to Thursday and increases since Friday. One can also observe that the minimum of average number curve for weekend is slightly larger than other days, which is consistent with Fig.~\ref{Fig:MMORPG:SP:Weekday}. Although this weekly pattern is significant, it is sufficient to consider only the intraday patterns for most quantitative analyses. These intraday and weekly patterns are quite similar for other virtual worlds. Because of the large number of independent game servers and the length of investigation period, our results can reveal reliably the common rules governing human dynamics. For instance, such circadian rhythms and weekly cycles are reported to generate universal macroscopic behaviors of humans \cite{Malmgren-Stouffer-Campanharo-Amaral-2009-Science,Aledavood-Lehmann-Saramaki-2015-FiP}.

\section{Probability distributions of $|\Delta N_i|$}
\label{S1:PDF}

The probability distribution of a random variable is of essential importance since it can fully determine the moments of the variable and may have a direct relationship to the multifractality of the time series \cite{Zhou-2012-CSF}. In this section, we will study the empirical probability distributions of $|\Delta N_i|$ of all the 95 virtual worlds.

\begin{figure}[!htb]
  \centering
  \includegraphics[width=8cm]{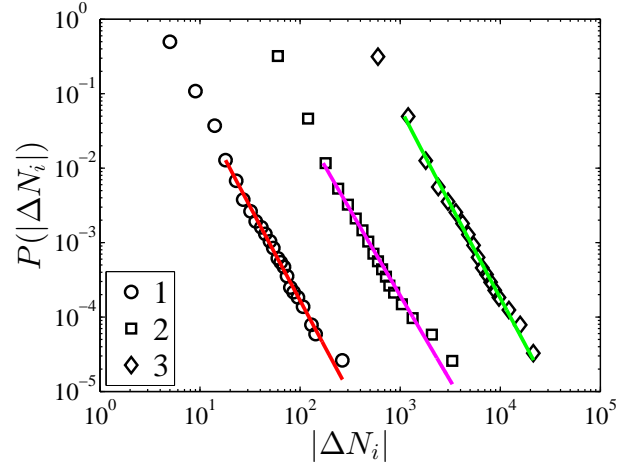}
  \caption{(color online) Empirical complementary cumulative distributions of the absolute activity fluctuations $|\Delta{N_i}|$ of online users in 1 min for three typical virtual worlds $i=1$, 2 and 3. The curves for $i=2$ and 3 have been shifted horizontally by 10 and 100 times for better visibility.}
  \label{Fig:MMORPG:PL:Fit:PDF}
\end{figure}

We computed the empirical probability distributions of $|\Delta N_i|$ for all the 95 virtual worlds and found that a large part of the distributions have power-law tails:
\begin{equation}
  p(|\Delta N_i|) = C_i |\Delta N_i|^{-(\beta_i+1)}
  \label{Eq:absDeltaN:PL}
\end{equation}
when $|\Delta N_i|\geq|\Delta N_i|_{\min}$, where $\beta_i$ is the power-law tail exponent, $|\Delta N_i|_{\min}$ is the lower bound of the scaling range of the power-law decay, and $C_i$ is the normalization factor.
Fig.~\ref{Fig:MMORPG:PL:Fit:PDF} illustrates the empirical complementary cumulative distributions $P(|\Delta N_i|)$ of $|\Delta N_i|$ for three virtual worlds. Evident power-law tails are observed. However, the objective detection and characterization of power-law tails is not straightforward due to the large fluctuations in the tails of the distributions. In particular, standard methods such as the least-squares regression on log-log scales are known to produce systematically biased estimates of the parameters for power-law distributions and thus should not be used under most circumstances \cite{Clauset-Shalizi-Newman-2009-SIAMR}.

To have a deeper understanding of the tail behavior, we need to conduct an objective analysis. Based on the Kolmogorov-Smirnov test, Clauset et al. proposed an efficient quantitative method to test if the tail has a power-law form and, if so, to estimate the power-law exponent $\beta_i$ for the data greater than or equal to a threshold $|\Delta N_i|_{\min}$ \cite{Clauset-Shalizi-Newman-2009-SIAMR}. We describe briefly the method, which has been extensively applied in many fields. Because the values of $|\Delta N_i|$ are positive integers, we mainly focus on the discrete case. Under the assumption of probability distribution form of $|\Delta N_i|$ described in Eq.~(\ref{Eq:absDeltaN:PL}), by calculating the normalizing constant, one finds that
\begin{equation}
 p(|\Delta{N_i}|)=C|\Delta N_i|^{-(\beta_i+1)}=\frac{|\Delta N_i|^{-(\beta_i+1)}}{\zeta(\beta_i+1,|\Delta N_i|_{\min})}
   \label{Eq:absDeltaN:PL:Final}
\end{equation}
where
\begin{equation}
  \zeta(\beta_i,|\Delta N_i|_{\min})=\sum_{0}^{\infty}(m+|\Delta N_i|_{\min})^{-\beta_i}
  \label{Eq:zeta:fun}
\end{equation}
is the generalized or Hurwitz zeta function and the variable $m$ in Eq.~(\ref{Eq:zeta:fun}) is the number of data points that $|\Delta N_i|\geq |\Delta N_i|_{\min}$. In addition, the complementary cumulative distribution is:
\begin{equation}
 P(|\Delta N_i|)=\frac{\zeta(\beta_i+1,|\Delta N_i|)}{\zeta(\beta_i+1,|\Delta N_i|_{\min})}.
\end{equation}

One can then determine the estimates of $|\Delta N_i|_{\min}$ and $\beta_i$ \cite{Clauset-Shalizi-Newman-2009-SIAMR}. The Kolmogorov-Smirnov statistic (KS) is defined as:
\begin{equation}\label{Eq:KS:Test}
  {\mathrm{KS}}=\max \limits_{|\Delta N_i|\geq |\Delta N_i|_{\min}}(|F(|\Delta N_i|)-F_{\mathrm{PL}}(|\Delta N_i|)|),
\end{equation}
\begin{table*}[ht]\addtolength{\tabcolsep}{2pt}
  \centering
  \caption{Characteristic parameters in the power-law distributions of $|\Delta N_i|$ for the 44 identified virtual worlds based on the Kolmogorov-Smirnov tests and the maximum likelihood estimation.}
  \label{TB:PL:Parameters}
  \resizebox{\textwidth}{!}{
   \begin{tabular}{*{17}{c}}
   \hline\hline
  No.&$|\Delta N_i|_{\min}$&$|\Delta N_i|_{\rm tail}$ &$\beta_i$ & $\sigma_{\beta_i}$ &$KS$&$p$-value&&No.&$|\Delta N_i|_{\min}$&$|\Delta N_i|_{\rm tail}$ &$\beta_i$ &$\sigma_{\beta_i}$ &$KS$&$p$-value \\
   \hline
  1&15&2893  &2.39 & 0.020 & 0.022 &0.42 &&23& 29&248   &2.20 &0.100&0.058 & 0.29\\
  2&17&1535  &2.28 & 0.124 & 0.019 &0.52 &&24& 31&218   &1.59 &0.031&0.068 & 0.56\\
  3& 8&13493 &2.41 & 0.005 & 0.023 &0.56 &&25& 21&200   &1.78 &0.309&0.047 & 0.60\\
  4&34&888   &2.06 & 0.221 & 0.041 &0.08 &&26& 24&326   &2.21 &0.137&0.031 & 0.41\\
  5&50&214   &1.66 & 0.111 & 0.041 &0.53 &&27& 11&9591  &2.42 &0.047&0.025 & 0.12\\
  6&37&396   &1.81 & 0.342 & 0.025 &0.55 &&28& 28&528   &1.67 &0.231&0.021 & 0.54\\
  7&50&189   &1.45 & 0.131 & 0.071 &0.14 &&29& 28&654   &2.08 &0.176&0.018 & 0.54\\
  8&25&1025  &1.94 & 0.068 & 0.017 &0.61 &&30& 48&220   &2.46 &0.112&0.044 & 0.79\\
  9&12&1189  &2.26 & 0.035 & 0.034 &0.18 &&31& 32&329   &1.86 &0.001&0.035 & 0.96\\
 10&16&3834  &2.40 & 0.053 & 0.017 &0.16 &&32& 25&1279  &2.35 &0.098&0.016 & 0.27\\
 11&12&5849  &2.42 & 0.019 & 0.012 &0.24 &&33& 24&500   &2.31 &0.025&0.028 & 0.35\\
 12&11&8468  &2.48 & 0.024 & 0.012 &0.15 &&34& 24&978   &2.35 &0.097&0.032 & 0.64\\
 13&11&12028 &2.15 & 0.044 & 0.015 &0.11 &&35& 43&194   &2.35 &0.179&0.033 & 0.40\\
 14&27&427   &1.97 & 0.053 & 0.013 &0.74 &&36&  7&21307 &2.50 &0.134&0.037 & 0.08\\
 15&12&5985  &2.45 & 0.039 & 0.008 &0.12 &&37&  6&20744 &2.44 &0.032&0.033 & 0.09\\
 16&29&472   &2.18 & 0.009 & 0.031 &0.43 &&38& 22&723   &2.16 &0.166&0.027 & 0.21\\
 17&29&668   &2.16 & 0.046 & 0.040 &0.38 &&39& 10&13440 &2.38 &0.038&0.027 & 0.07\\
 18&45&339   &2.07 & 0.147 & 0.018 &0.83 &&40& 26&552   &2.04 &0.218&0.023 & 0.42\\
 19&32&985   &2.20 & 0.140 & 0.025 &0.22 &&41&  7&20614 &2.47 &0.017&0.020 & 0.14\\
 20&45&268   &1.92 & 0.281 & 0.042 &0.13 &&42& 58&130   &2.21 &0.105&0.109 & 0.07\\
 21&10&15661 &2.19 & 0.073 & 0.015 &0.14 &&43& 58&117   &2.25 &0.061&0.117 & 0.07\\
 22&27&456   &2.06 & 0.216 & 0.027 &0.24 &&44& 35&227   &1.74 &0.272&0.033 & 0.31\\
   \hline\hline
   \end{tabular}}
\end{table*}where $F(|\Delta N_i|)$ is the cumulative distribution of the absolute fluctuations of online user activities and $F_{\mathrm{PL}}(|\Delta N_i|)$ is the cumulative distribution of the best power-law fit. The lower bound $|\Delta N_i|_{\min}$ is determined by minimizing the KS statistic. Then the power-law tail exponent $\beta_i$ of the data in the range $|\Delta N_i|\geq |\Delta N_i|_{\min}$ can be estimated using the maximum likelihood estimation (MLE) method as follows,
\begin{equation}\label{Eq:Discrete:Beta}
  \beta_i \simeq m\left[\sum_{i=1}^{m}\mathrm{ln}\frac{|\Delta N_i|}{|\Delta N_i|_{\min}-\frac{1}{2}}\right]^{-1}.
\end{equation}
The standard error $\sigma_{\beta_i}$ of the power-law exponent $\beta_i$ is derived from a quadratic approximation to the log-likelihood at its maximum, which reads
\begin{equation}\label{Eq:Discrete:Sigma}
  \sigma_{\beta_i}=\frac{1}{\sqrt{m\left[\frac{\zeta''(\beta_i+1,|\Delta N_i|_{\min})}{\zeta(\beta_i+1,|\Delta N_i|_{\min})}
  -\left[\frac{\zeta'(\beta_i+1,|\Delta N_i|_{\min})}{\zeta(\beta_i+1,|\Delta N_i|_{\min})}\right]^{2}\right]}}.
\end{equation}

%

Following Clauset et al. \cite{Clauset-Shalizi-Newman-2009-SIAMR}, to check whether the power-law tail is a plausible fit to the absolute fluctuations of virtual world activities, we perform the bootstrap test. In doing so, we generate 2500 realizations of power-law distributed synthetic data sets with the scaling parameter $\beta_i$ and the lower bound $|\Delta N_i|_{\min}$ equal to those of the distribution that best fits the observed data. Note that, if we wish the $p$-values to be accurate within about $\epsilon$ of the true value, we should generate at least $\frac{1}{4}\epsilon^{-2}$ synthetic data sets \cite{Clauset-Shalizi-Newman-2009-SIAMR}. Thus, if we wish the $p$-value to be accurate to about 2 decimal digits, we would choose $\epsilon=0.01$, which implies that we should generate about 2500 synthetic sets. We fit each synthetic data set individually to its own power-law model and calculate the statistic ${\mathrm{KS}}_{\mathrm{sim}}$ for each realization relative to its own model, which is as follows:
 \begin{equation}\label{Eq:KS:Test:sim}
  {\mathrm{KS}}_{\mathrm{sim}}=\max(|F_{\mathrm{sim}}-F_{\mathrm{PL}}|),
\end{equation}
where $F_{\mathrm{sim}}$ is the cumulative distribution of the synthetic realization. We calculate the fraction of resulting simulation statistics being larger than the value of the empirical data, that is,
\begin{equation}\label{Eq:PL:p:value}
  p-\mathrm{value}=\frac{\#({\mathrm{KS}}_{\mathrm{sim}}>{\mathrm{KS}})}{L_{\mathrm{sim}}}
\end{equation}
where the numerator is the number of realizations with ${\mathrm{KS}}_{\mathrm{sim}}>$KS and $L_{\mathrm{sim}} =2500$ is the number of synthetic realizations.

Applying these approaches to the absolute fluctuations of virtual world activities, we identify 44 cases out of the 95 virtual worlds that have power-law tails, in which the $p$-values are greater than 5\%. The determined  characteristic parameters $|\Delta N_i|_{\min}$, $|\Delta N_i|_{\rm tail}$, $\beta_i$ and $\sigma_{\beta_i}$ are presented in Table~\ref{TB:PL:Parameters}. We find that the power-law tail exponent mainly concentrates in the range $[2.2,2.5]$, while the minimum threshold $|\Delta N_i|_{\min}$ mainly concentrates in the range $[6,14]$ and $[22,38]$ and the number of observations in the power-law part of the distribution $|\Delta N_i|_{\rm tail}$ fluctuates a lot. Furthermore, Table~\ref{TB:PL:Parameters} shows that there are only six $p$-values less than 0.1 and most $p$-values are greater than 0.2. We also find that the tail exponent $\beta_i$ is small if $|\Delta N_i|_{\min}$ is large. A simple linear regression shows that
\begin{equation}
  \beta_i = a_0 + a_1 |\Delta N_i|_{\min}
\end{equation}
where $a_0=2.401$ and $a_1=-0.0097$ and the adjusted $R$-square is 0.24. The estimated values of the two coefficients $a_0$ and $a_1$ are significantly different from 0 with the $p$-values less than 0.1\%.

\section{Multifractal analysis}
\label{S1:MF}

\subsection{Partition function approach}


We apply the partition function approach to unveil the multifractal nature of the fluctuations of online user activities. Denote the 1-min absolute fluctuation time series of online users as $\{|\Delta N_i(t)|:t=1,2,...,T \}$. The time series is covered by $V$ boxes with equal sizes, where $V=[T/s]$. On each box $B(v,s)=[(v-1)s+1,vs]$ with $v=1,2,\cdots,V$, we define a quantity $u$ as follows:
\begin{equation}
  u(v;s)=u(B(v,s))=\sum_{t=1}^s|\Delta{N_i}((v-1)s+t)|.
\end{equation}
The box sizes $s$ are chosen such that $V=[T/s]=T/s$. The measure $\mu$ on each box is constructed as follows:
\begin{equation}
  \mu(v;s)= {u(v;s)}/{U},
  \label{Eq:u}
\end{equation}
where $U=\sum_{v=1}^Vu(v;s)=\sum_{t=1}^T|\Delta{N_i}(t)|$. We can calculate the partition function $\chi_{q}$ \cite{Halsey-Jensen-Kadanoff-Procaccia-Shraiman-1986-PRA}:
\begin{equation}\label{Eq:PF}
  \chi_{q}(s) = \sum_{v=1}^V[\mu(v;s)]^q,
\end{equation}
and expect it to scale as
\begin{equation}
  \chi_{q}(s) \sim s^{\tau(q)},
\end{equation}
where the exponent $\tau(q)$ is the mass scaling exponent function. The local singularity strength $\alpha$ of the measure $\mu$ and its spectrum $f(\alpha)$ are related to $\tau(q)$ through the Legendre transformation \cite{Halsey-Jensen-Kadanoff-Procaccia-Shraiman-1986-PRA}:
\begin{equation}
    \left\{
    \begin{array}{ll}
        \alpha(q)={\rm{d}}\tau(q)/{\rm{d}}q\\
        f(q)=q{\alpha}-{\tau}(q)
    \end{array}
    \right..
\label{Eq:f:alpha:tau}
\end{equation}

When $\mu(v;s)\ll1$ and $q\gg1$, the corresponding value of the partition function $\chi$ will be too small such that the computer becomes ``out of memory''. To overcome this problem, we can calculate the logarithm of the partition function ln$\chi_{q}(s)$ rather than the partition function itself, together with a simple manipulation of Eqs.~(\ref{Eq:u}) and (\ref{Eq:PF}), which results in the following formula \cite{Jiang-Zhou-2008b-PA}:
\begin{eqnarray}
  \ln\chi_{q}(s)&=&\ln\sum_{v=1}^{V}\left[\frac{u(v;s)}{u_{\max}}\times \frac{u_{\max}}{U}\right]^q \nonumber\\
  &=&\ln\sum_{v=1}^{V}\left[\frac{u(v;s)}{u_{\max}}\right]^q +q\ln\left[\frac{u_{\max}}{U}\right],
\end{eqnarray}
where $u_{\max}=\max \limits_{v}\{u(v;s)\}$ is the maximum of $u(v;s)$ for $v=1,2,...,V$.

\subsection{Empirical results}

\begin{figure}[h]
  \centering
  \includegraphics[width=4cm]{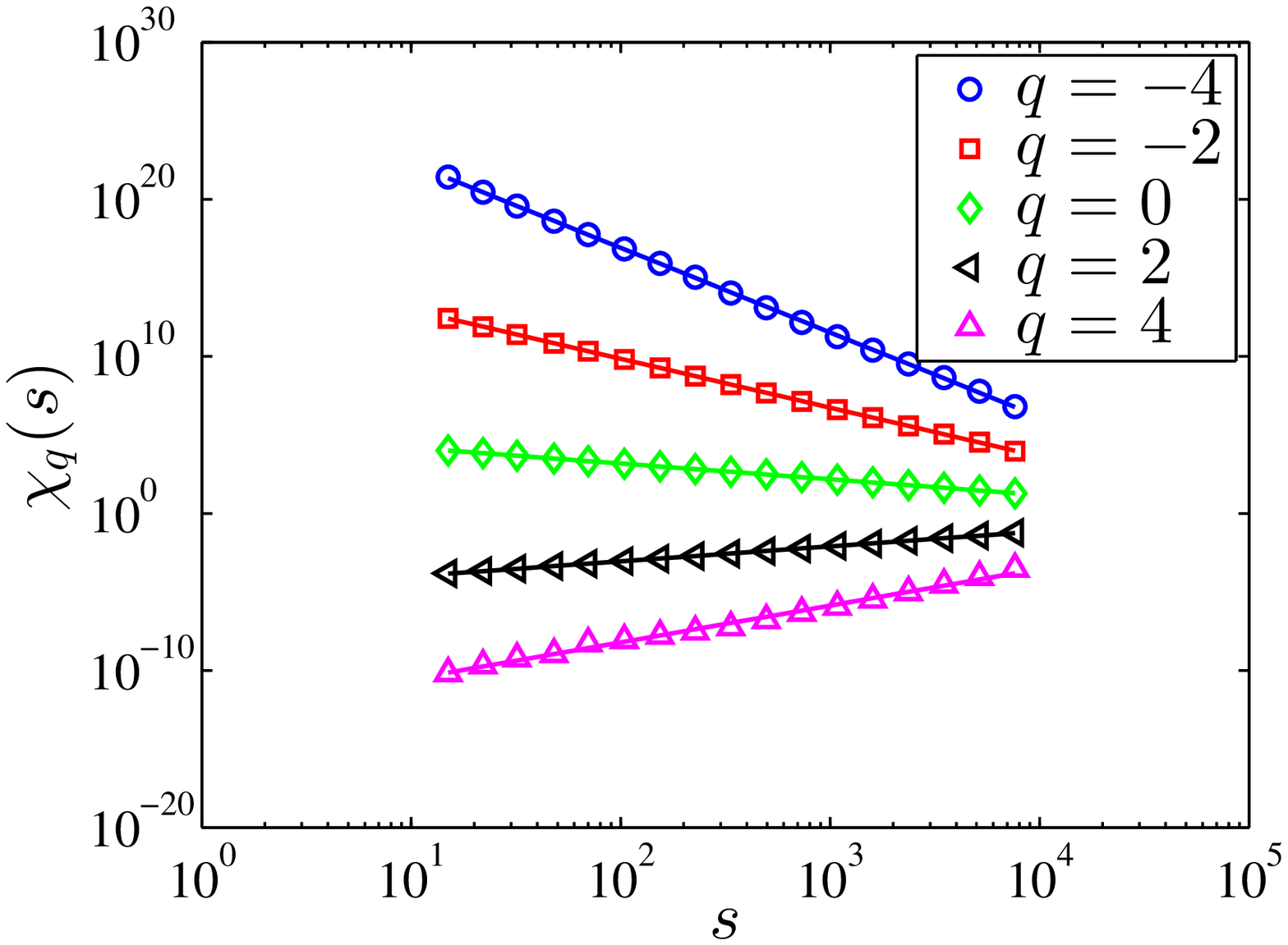}
  \includegraphics[width=4cm]{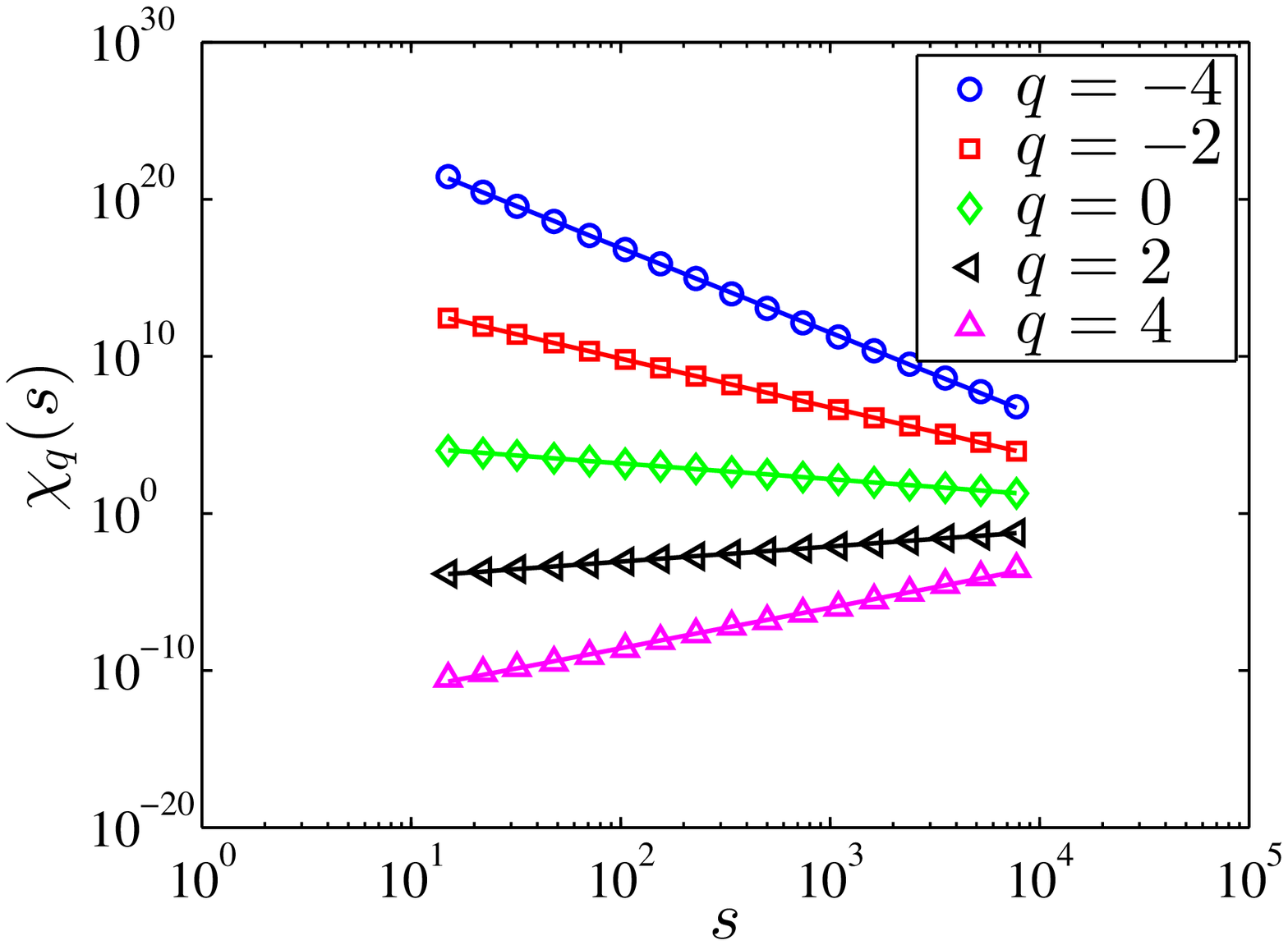}\\
  \includegraphics[width=4cm]{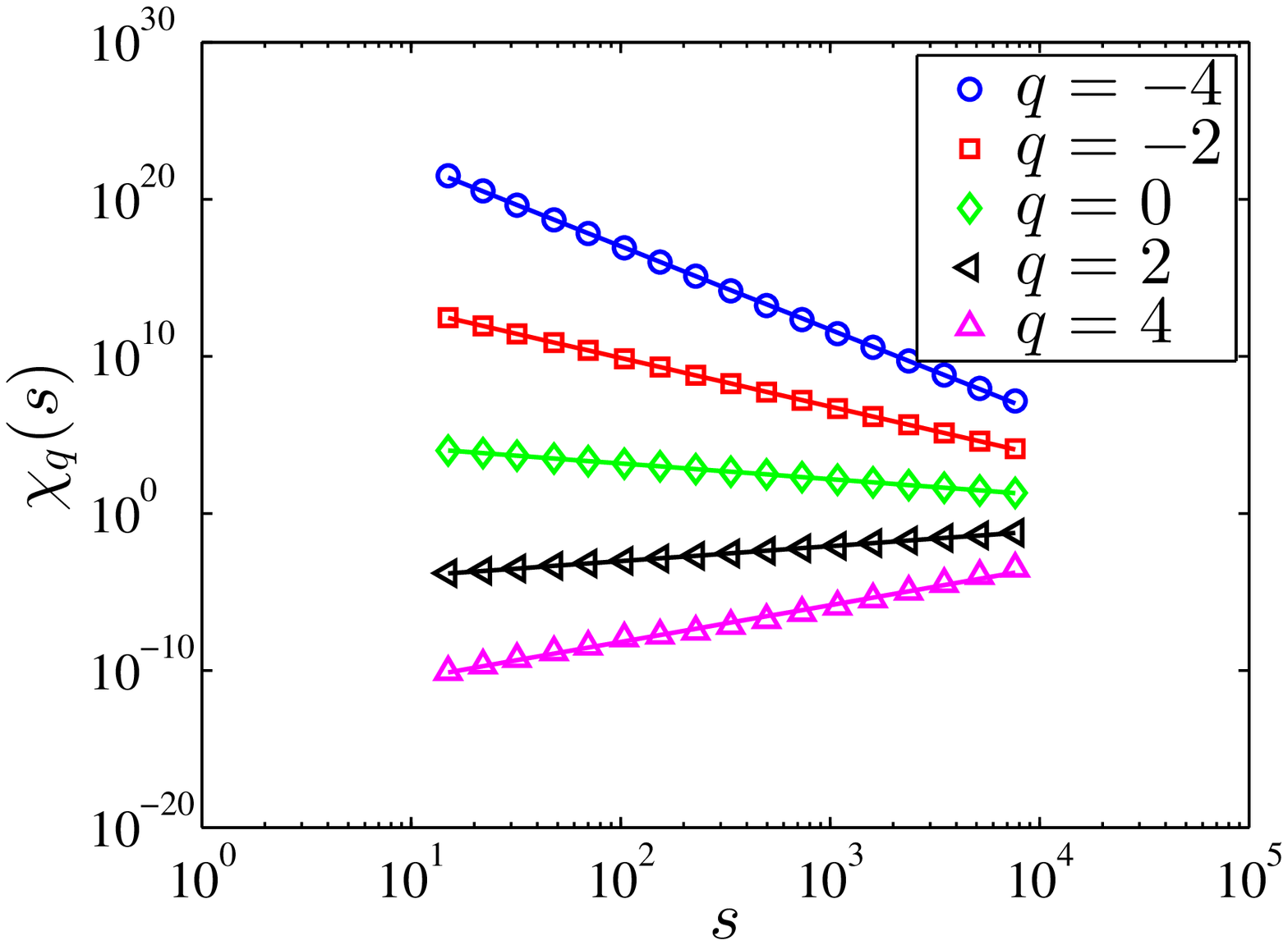}
  \includegraphics[width=4cm]{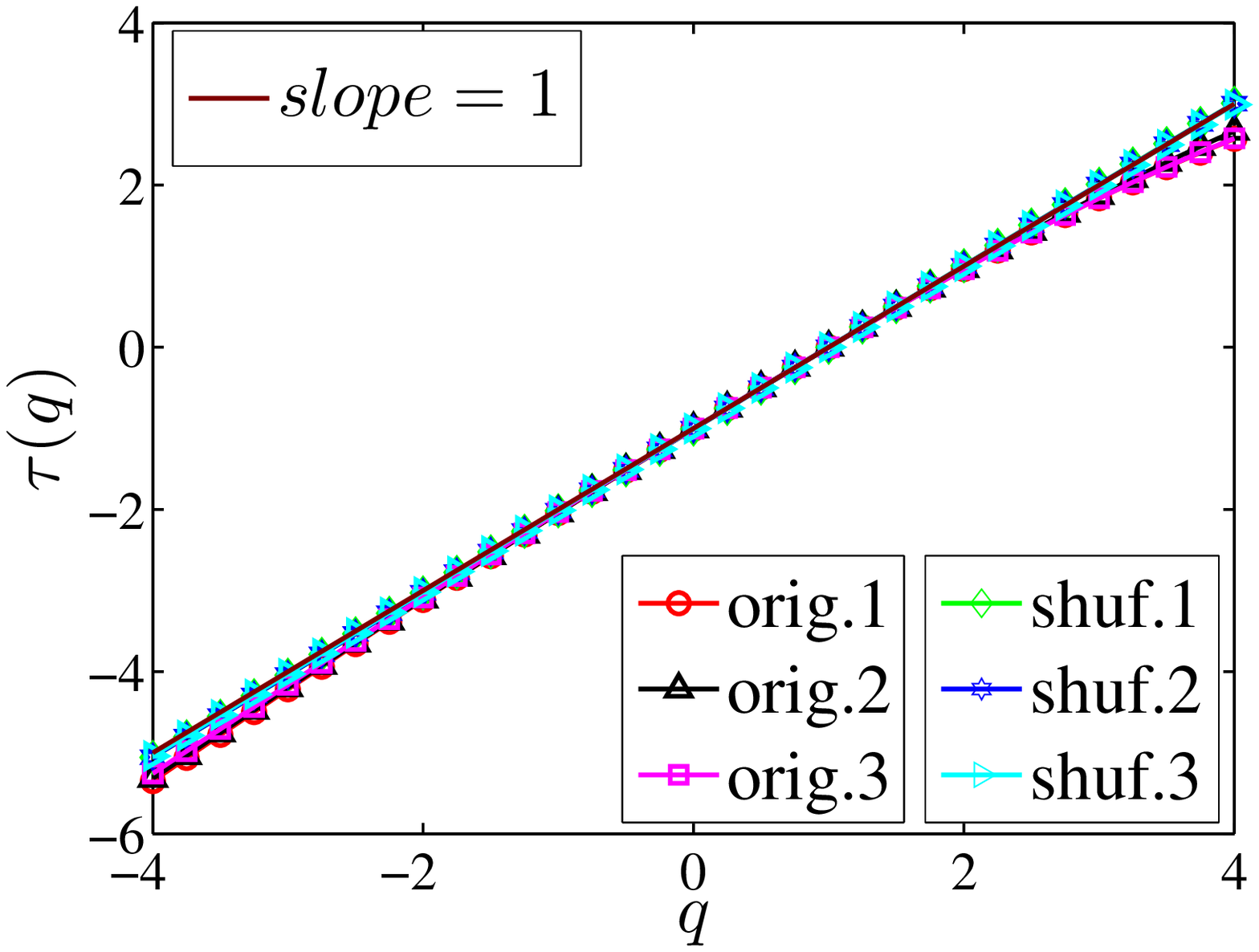}\\
  \includegraphics[width=4cm]{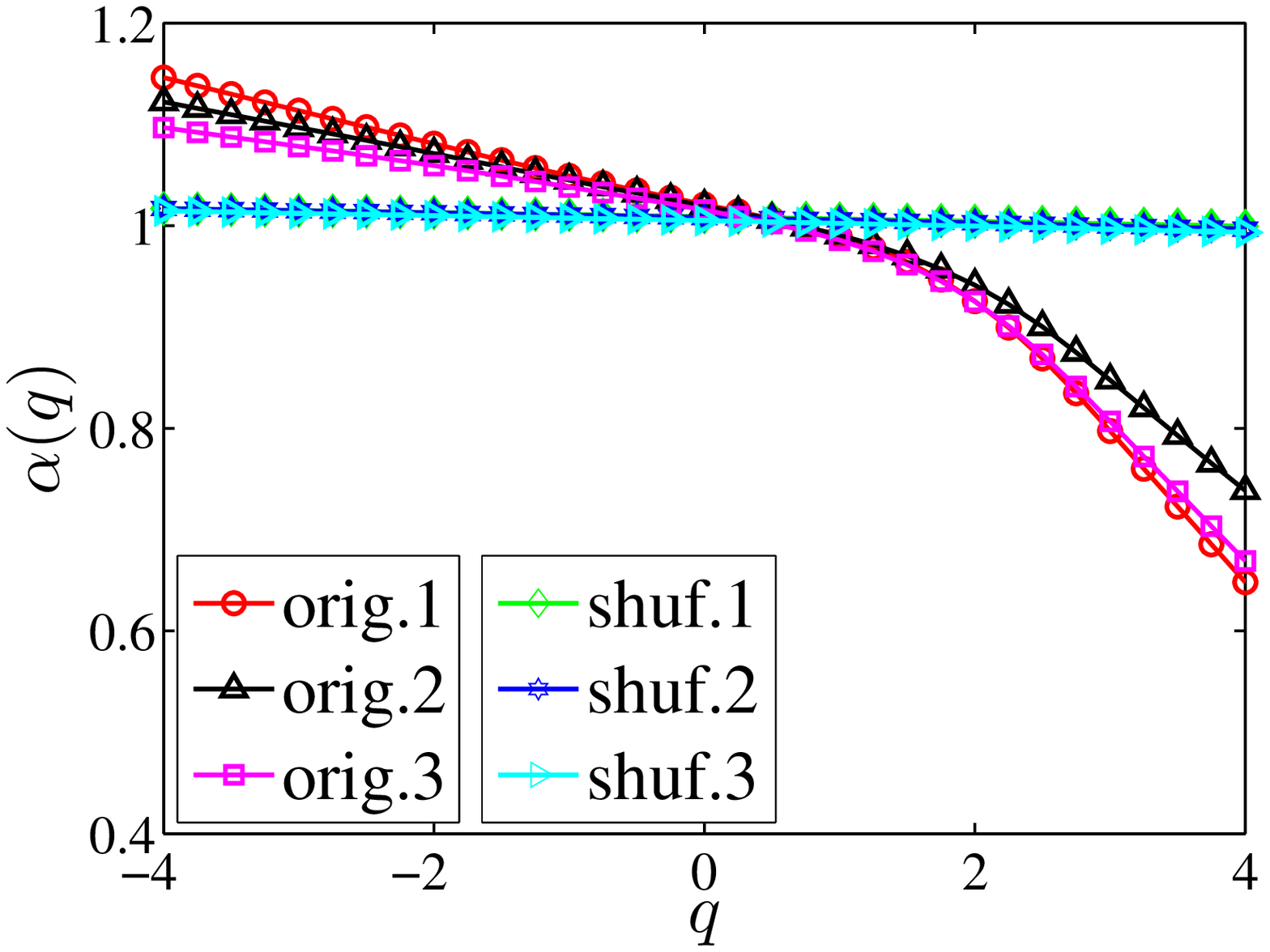}
  \includegraphics[width=4cm]{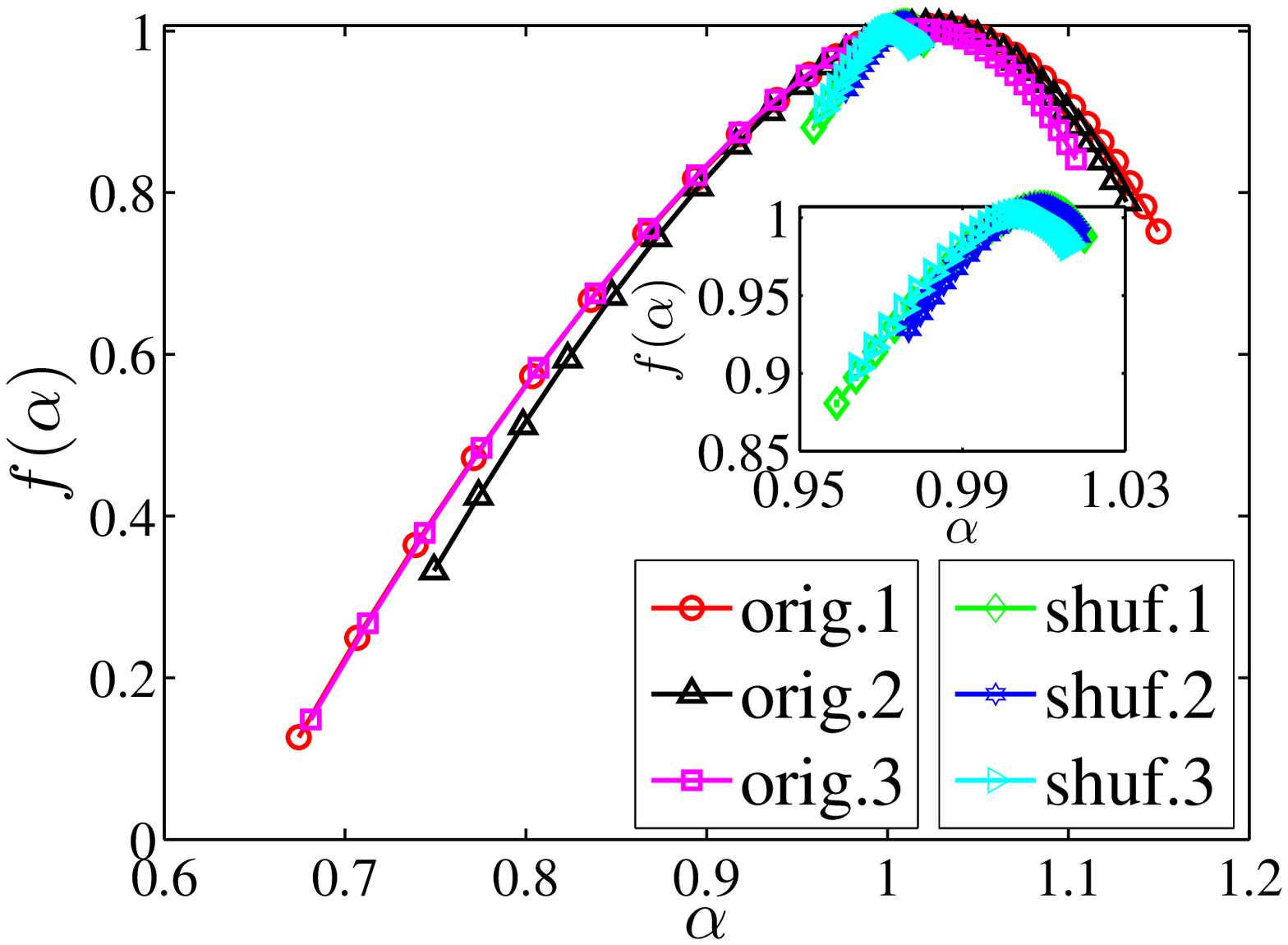}
  \vskip    -9.2cm   \hskip     -8.1cm {\textsf{(a)}}
  \vskip    -0.3cm   \hskip     0.3cm {\textsf{(b)}}
  \vskip    2.6cm   \hskip       -8.2cm {\textsf{(c)}}
  \vskip    -0.3cm   \hskip     0.3cm {\textsf{(d)}}
  \vskip    2.5cm   \hskip     -8.2cm {\textsf{(e)}}
  \vskip    -0.3cm   \hskip     0.3cm {\textsf{(f)}}
  \vskip    2.8cm
  \caption{(color online) Multifractal analysis of 1-min absolute fluctuations $|\Delta{N}_i(t)|$ of online users. (a-c) Power-law dependence of $\chi_{q}(s)$ on the box size $s$ for different $q$ of original $|\Delta{N}_i(t)|$. (d) The mass exponent functions $\tau(q)$ of the original $|\Delta{N}_i(t)|$ and the corresponding shuffled series. (e) The singularity strength functions $\alpha(q)$ of original $|\Delta{N}_i(t)|$ and its shuffled series. (f) The multifractal singularity spectra $f(\alpha)$ of original $|\Delta{N}_i(t)|$ and its shuffled series. The inset is the amplification of the main plot around $(a,f)=(1,1)$ which only amplify the shuffled series.}
  \label{Fig:MMORPG:Multifractal}
\end{figure}

Plots (a-c) of Fig.~\ref{Fig:MMORPG:Multifractal} show the dependence of the partition function $\chi_{q}(s)$ on the box size $s$ for different values of $q$ in log-log coordinates for three virtual worlds. We find that the partition functions $\chi_{q}(s)$ scale as excellent power laws with respect to $s$, with the scaling range spanning about three orders of magnitude. The mass exponents $\tau(q)$ are estimated by the slopes of the linear fits to $\ln\chi_{q}(s)$ with respect to $\ln{s}$ for different values of $q$, which are shown in Fig.~\ref{Fig:MMORPG:Multifractal}(d). One observes that there is no evident linear relationship between $\tau(q)$ and $q$ for all the three examples, which is concluded due to the deviation of the $\tau(q)$ curves of the origin time series from the counterpart $\tau(q)$ curves of the shuffled time series that are linear. The nonlinearity of the mass exponent functions indicates that the online user fluctuations exhibit multifractal nature.

According to the Legendre transformation, we numerically determine the singularity strength functions $\alpha(q)$, which are the first-order derivatives of the corresponding $\tau(q)$ functions, and the multifractal functions $f(\alpha)$. The results for $\alpha(q)$ are illustrated in Fig.~\ref{Fig:MMORPG:Multifractal}(e). One finds that the $\alpha(q)$ function of the second time series is greater than the rest two functions when $q\geq2$, which corresponds to the difference in the multifractal singularity spectra $f(\alpha)$ shown in Fig.~\ref{Fig:MMORPG:Multifractal}(f). The nonlinearity of $\alpha(q)$ and the broad spectrum of $f(\alpha)$ are both hallmarks of multifractality. We note that online user fluctuations in other virtual worlds also exhibit evident multifractal nature.

\subsection{Multifractality and system performance}

We now turn to investigate if there are any relationships between the multifractal nature of online user fluctuations and the performance of virtual worlds. The most often used conventional measure for quantifying the degree of multifractality is the width of singularity spectrum, which can be calculated as follows,
\begin{equation}\label{Eq:Delta:alpha}
 \Delta\alpha_i=\alpha_{i,\max}-\alpha_{i,\min}
\end{equation}
although alternative measures are also used in some cases \cite{Zunino-Tabak-Figliola-Perez-Garavaglia-Rosso-2008-PA,deSouza-Queiros-2009-CSF}. The wider is the singularity spectrum $\Delta\alpha$, the stronger is the multifractality in the time series. We find that the width of singularity spectrum $\Delta\alpha$ concentrates in the range $[0.3,0.6]$. This observation confirms that all the 95 time series associated with the 95 virtual worlds exhibit multifractal nature \cite{Zhou-2012-CSF}.

Further, we adopt two measures to quantify the performance of the virtual worlds. The first performance measure is the time elapsed for a new world to reach its maximum activity $n_{i,\max}$, denoted by $t_{i,\max}$. There are 53 (55.79\%) virtual worlds whose $t_{\max}$ values are less than one day and the mean is 2.697 days, indicating a very fast growth of users after a new world is created in a server. However, there are also a small amount of virtual worlds whose $t_{\max}$ values are greater than 5 days. The second performance measure is the relative maximum activity $n_{i,\max}$ for each virtual world, as defined in Eq.~(\ref{Eq:ni:max}).

We propose the following linear relationship
\begin{equation}
  \Delta\alpha_i =a_0 + a_n n_{i,\max} + a_t t_{i,\max},
  \label{Eq:Dalpha:nmax:tmax}
\end{equation}
where $a_0$, $a_n$ and $a_t$ are the regression coefficients. We perform least-squares robust linear regressions of Eq.~(\ref{Eq:Dalpha:nmax:tmax}) for three samples: all the 95 virtual worlds, the 44 virtual worlds whose activity fluctuation distributions have power-law tails, and the rest 51 virtual worlds whose activity fluctuation distributions do not have power-law tails. The estimated coefficients $a_0$, $a_n$ and $a_t$ are presented in Table \ref{TB:Dalpha:nmax:tmax}, together with the associated $p$-values. We find that only the relation for the 44 virtual worlds with power-law tail distributions in the activity fluctuations is statistically significant. For this sample, the $F$-statistic is 3.868, the $p$-value is 0.029, the $R^2$ is 0.159, and the adjusted $R^2$ is 0.118. Furthermore, $n_{i,\max}$ is different from 0 at the 7\% significance level and $t_{i,\max}$ is different from 0 at the 1\% significance level.

\begin{table}[ht]\addtolength{\tabcolsep}{-2pt}
  \centering
  \caption{Testing the possible dependence of the width of singularity strength $\Delta\alpha_i$ on the maximum relative activity $n_{i,\max}$ and the time $t_{i,\max}$ to the maximum relative activity using a linear equation (\ref{Eq:Dalpha:nmax:tmax}). The sample ``All'' contains all the 95 virtual worlds under investigation. The sample ``PL'' contains the 44 virtual worlds with power-law activity fluctuation distributions. The sample ``Non-PL'' contains the 51 virtual worlds with non-power-law distributions in the activity fluctuations.}
  \label{TB:Dalpha:nmax:tmax}
   \begin{tabular}{ccccccccccccc}
   \hline\hline
  Sample & $a_0$ & $p_0$ & $a_n$ & $p_n$ & $a_t$ & $p_t$ & Adj-$R^2$\\
  \hline
  All    & 0.523 & 0 & -0.070 & 0.101 & -0.0014 & 0.367 & 0.009\\
  PL     & 0.560 & 0 & -0.091 & 0.070 & -0.0065 & 0.010 & 0.118\\
  Non-PL & 0.501 & 0 & -0.058 & 0.383 & +0.0020 & 0.303 & 0.019\\
   \hline\hline
   \end{tabular}
\end{table}

Table \ref{TB:Dalpha:nmax:tmax} shows that, for the 44 virtual worlds with power-law activity fluctuation distributions, the activity fluctuations exhibit stronger multifractality if the users grow faster after the virtual world is set up. In addition, there is a weak effect that virtual worlds with high user activities might have weaker multifractality. These two effects are indeed consistent with each other and an intuitive interpretation is store. When there are less active users and the virtual system reaches maturity faster, the user activity may fluctuate relatively severely, which results in larger intermittence and stronger multifractality.

\subsection{The components of multifractality in $|\Delta{N}_i(t)|$ series}

Generally speaking, understanding the components of multifractality is an important and subtle issue. There are a wealth of studies showing that multifractal nature is usually attributed to the influence of temporal structure (linear correlation and nonlinearity) and fat-tailedness in the probability distribution \cite{Kantelhardt-Zschiegner-KoscielnyBunde-Havlin-Bunde-Stanley-2002-PA,Zhou-2009-EPL,Zhou-2012-CSF}. In addition, one can quantitatively determine the contribution
of the temporal structure and the fat-tail components through the singularity width $\Delta\alpha$ of the multifractal spectrum.

To understand the impact of the temporal structure, we shuffle the $|\Delta{N}_i(t)|$ series 100 times and determine their singularity spectra \cite{Theiler-Eubank-Longtin-Galdrikian-Farmer-1992-PD,Schreiber-Schmitz-1996-PRL,Schreiber-Schmitz-2000-PD}. For each point on the $\tau(q)$, $\alpha(q)$ and $f(\alpha)$ curve of the shuffled data, $\tau(q)$, $\alpha(q)$ and $f(\alpha)$ are the arithmetic averages of the respective 100 values of the shuffled data. And the almost invisible error bar is the corresponding standard deviation which is extremely close to zero. The results are depicted in Fig.~\ref{Fig:MMORPG:Multifractal}. One can find that the singularity width of shuffled data shrinks remarkably by observing Fig.~\ref{Fig:MMORPG:Multifractal}(f). These observations imply that the temporal structure (linear correlation and nonlinearity) of the $|\Delta{N}_i(t)|$ series has a crucial impact on the singularity width $\Delta\alpha$.

In addition, given that some $|\Delta{N}_i(t)|$ time series have broad distributions of fluctuations which can be observed in Fig.~\ref{Fig:MMORPG:dn:Evolution}, it is natural to conjecture whether the large fluctuations have remarkable contribution to the observed multifractality of $|\Delta{N}_i(t)|$. In financial markets, the null hypothesis that the reported multifractal nature stems from the large price fluctuations cannot be rejected \cite{Lux-2004-IJMPC}.

\begin{figure}[!htb]
  \centering
  \includegraphics[width=8cm]{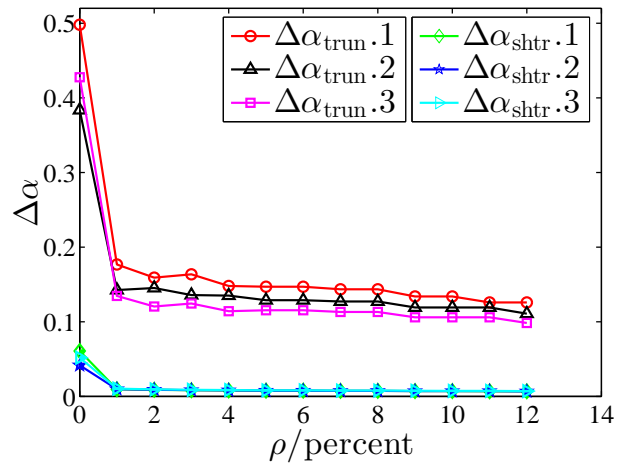}
  \caption{(color online) Dependence of the singularity width $\Delta\alpha$ on the fixed $\rho\%$ for the truncated time series from which the largest $\rho\%$ of absolute activity fluctuations $|\Delta{N_i}|$ have been deleted $(\Delta\alpha_{\rm{trun}})$ and for the shuffled truncated data $(\Delta\alpha_{\rm{shtr}})$.}
  \label{Fig:MMORPG:Truncated:Dalpha}
\end{figure}

To investigate the impact of broad distributions of fluctuations on the singularity width $\Delta\alpha$ of the multifractal spectrum, one can either remove large values \cite{Oh-Eom-Havlin-Jung-Wang-Stanley-Kim-2012-EPJB} or generate surrogate data \cite{Theiler-Eubank-Longtin-Galdrikian-Farmer-1992-PD,Schreiber-Schmitz-1996-PRL,Schreiber-Schmitz-2000-PD,Zhou-2009-EPL,Zhou-2012-CSF}. We have followed the first idea and constructed the truncated time series by eliminating the largest $\rho\%$ of $|\Delta{N}_i(t)|$. For convenience, the resulting data are termed as truncated data. We have generated 13 truncated data sets with the truncation percentage $\rho$ spanning from 0 to 12 with a spacing step of 1. The dependence of the singularity width $\Delta\alpha_{\rm{trun}}$ on $\rho$ is illustrated in Fig.~\ref{Fig:MMORPG:Truncated:Dalpha}. The singularity width $\Delta\alpha$ drops dramatically after removing the top 1\% data and then decreases smoothly with increasing $\rho$, which remains at a relatively high level. For each value of $\rho$, the truncated data are shuffled to generate 100 shuffled truncated data sets. For each shuffled truncated data set, we determine its average singularity width $\Delta\alpha_{\rm{shtr}}$, which is very close to 0. The results are also presented in Fig.~\ref{Fig:MMORPG:Truncated:Dalpha} for comparison. These analyses indicate that large values in the $|\Delta{N}_i(t)|$ series have a significant impact on $\Delta\alpha$, which is however not the unique influencing factor. The temporal structure of the truncated series is also a strong factor on its multifractality. What's more, we have investigated the components of multifractality in other virtual worlds and observed similar results.

\section{Conclusion}
\label{S1:Conclusion}

In summary, we first characterized the basic properties of the time series of 1-min instant number of online users or user activity $N_i(t)$ of 95 independent virtual worlds of an MMORPG inhabited in different servers. We confirmed the presence of intraday patterns and weekly patterns in user activities, which are common traits in human dynamics due to circadian rhythms and weekly cycles \cite{Malmgren-Stouffer-Campanharo-Amaral-2009-Science,Jiang-Ren-Gu-Tan-Zhou-2010-PA,Jo-Karsai-Kertesz-Kaski-2012-NJP,Jiang-Xie-Li-Podobnik-Zhou-Stanley-2013-PNAS}. Based on an effective approach proposed by Clauset et al. \cite{Clauset-Shalizi-Newman-2009-SIAMR}, we identified that there are 44 time series of absolute activity fluctuations $|\Delta N_i(t)|$ following power-law distributions in the tails. We estimated that the power-law tail exponent mainly concentrates in the range $[2.2,2.5]$.

In order to detect the multifractality in absolute activity fluctuations $|\Delta N_i(t)|$ and its relation to the performance of virtual worlds, we have performed multifractal analysis of $|\Delta N_i(t)|$, based on the partition function approach. We found that all the 95 time series exhibit evident multifractal nature. We further found the width of singularity strength $\Delta\alpha$ is negatively correlated with the maximum activity $n_{i,\max}$ at the significance level of 7\% and with the time to the maximum relative activity $t_{\max}$  at the significance level of 1\% for the 44 virtual worlds with power-law tailed distributions in their absolute activity fluctuations. Our findings show that the strength of multifractality in the absolute user activity fluctuations reflects to some extent the self-organized behavior of complex socioeconomic systems and is able to quantify their evolution behaviors.

Additionally, we find both temporal structure and large fluctuations contribute to the multifractality of the $|\Delta{N}_i(t)|$ series, while the temporal structure plays a major role. However, we should note that these analyses unveil only the mechanical sources of multifractality originated from other statistical properties of the time series, which does not provide a physical mechanism. Further studies are required to understand the mechanism that causes the emergence of multifractality in the fluctuations of user activities by agent-based modelling.

\section*{Acknowledgement}

We are grateful to the anonymous reviewers for their insightful comments and suggestions. Errors are ours. This work was supported by the National Natural Science Foundation of China (11375064, 11505063, 11605062), the Postdoctoral Science Foundation of China (2016M600291), the Shanghai Chenguang Program (15CG29),  and the Fundamental Research Funds for the Central Universities (222201718006).


\begin{thebibliography}{10}
\expandafter\ifx\csname url\endcsname\relax
  \def\url#1{\texttt{#1}}\fi
\expandafter\ifx\csname urlprefix\endcsname\relax\def\urlprefix{URL }\fi
\expandafter\ifx\csname href\endcsname\relax
  \def\href#1#2{#2} \def\path#1{#1}\fi

\bibitem{Bainbridge-2007-Science}
W.~S. Bainbridge, {The scientific research potential of virtual worlds},
  Science 317~(5837) (2007) 472--476.
\newblock \href {http://dx.doi.org/10.1126/science.1146930}
  {\path{doi:10.1126/science.1146930}}.

\bibitem{Grabowski-Kosinski-2008-APPA}
A.~Grabowski, R.~Kosi{\'n}ski, {The SIRS model of epidemic spreading in virtual
  society}, Acta Phys. Pol. A 114 (2008) 589--596.

\bibitem{Tseng-Li-Wang-2010-EPJB}
J.-J. Tseng, S.-P. Li, S.-C. Wang, {Experimental evidence for the interplay
  between individual wealth and transaction network}, Eur. Phys. J. B 73 (2010)
  69--74.
\newblock \href {http://dx.doi.org/10.1140/epjb/e2009-00424-8}
  {\path{doi:10.1140/epjb/e2009-00424-8}}.

\bibitem{Castronova-2001-WP}
E.~Castronova, {Virtual worlds: A first-hand account of market and society on
  the cyberian frontier}, available at SSRN: http://ssrn.com/abstract=294828
  (2001).

\bibitem{Grabowski-2007-PA}
A.~Grabowski, {Interpersonal interactions and human dynamics in a large social
  network}, Physica A 385 (2007) 363--369.
\newblock \href {http://dx.doi.org/10.1016/j.physa.2007.06.005}
  {\path{doi:10.1016/j.physa.2007.06.005}}.

\bibitem{Grabowski-2009-PA}
A.~Grabowski, {Opinion formation in a social network: The role of human
  activity}, Physica A 388 (2009) 961--966.
\newblock \href {http://dx.doi.org/10.1016/j.physa.2008.11.036}
  {\path{doi:10.1016/j.physa.2008.11.036}}.

\bibitem{Grabowski-Kosinski-2008-APPB}
A.~Grabowski, R.~Kosi{\'n}ski, {Mixing patterns in a large social network},
  Acta Phys. Pol. B 39 (2008) 1291--1300.

\bibitem{Grabowski-Kruszewska-2007-IJMPC}
A.~Grabowski, N.~Kruszewska, {Experimental study of the structure of a social
  network and human dynamics in a virtual society}, Int. J. Mod. Phys. C 18
  (2007) 1527--1535.
\newblock \href {http://dx.doi.org/10.1142/S0129183107011480}
  {\path{doi:10.1142/S0129183107011480}}.

\bibitem{Grabowski-Kruszewska-Kosinski-2008-EPJB}
A.~Grabowski, N.~Kruszewska, R.~A. Kosi{\'n}ski, {Properties of on-line social
  systems}, Eur. Phys. J. B 66 (2008) 107--113.
\newblock \href {http://dx.doi.org/10.1140/epjb/e2008-00379-2}
  {\path{doi:10.1140/epjb/e2008-00379-2}}.

\bibitem{Grabowski-Kruszewska-Kosinski-2008-PRE}
A.~Grabowski, N.~Kruszewska, R.~A. Kosi{\'n}ski, {Dynamic phenomena and human
  activity in an artificial society}, Phys. Rev. E 78 (2008) 066110.
\newblock \href {http://dx.doi.org/10.1103/PhysRevE.78.066110}
  {\path{doi:10.1103/PhysRevE.78.066110}}.

\bibitem{Xie-Li-Jiang-Tan-Podobnik-Zhou-Stanley-2016-SR}
W.-J. Xie, M.-X. Li, Z.-Q. Jiang, Q.-Z. Tan, B.~Podobnik, W.-X. Zhou, H.~E.
  Stanley, {Skill complementarity enhances heterophily in collaboration
  networks}, Sci. Rep. 6~(1) (2016) 18727.
\newblock \href {http://dx.doi.org/10.1038/srep18727}
  {\path{doi:10.1038/srep18727}}.

\bibitem{Matsuda-2003-Presence}
K.~Matsuda, {Can we sell a virtual object in a virtual society?}, Presence 12
  (2003) 581--598.
\newblock \href {http://dx.doi.org/10.1162/105474603322955897}
  {\path{doi:10.1162/105474603322955897}}.

\bibitem{Castronova-2005-HBR}
E.~Castronova, {Real products in imaginary worlds}, Harward Buss. Rev. 83~(5)
  (2005) 20--22.

\bibitem{Hemp-2006-HBR}
P.~Hemp, {Avatar-baed marketing}, Harward Buss. Rev. 84~(6) (2006) 48--57.

\bibitem{Papagiannidis-Bourlakis-Li-2008-TFSC}
S.~Papagiannidis, M.~Bourlakis, F.~Li, {Making real money in virtual worlds:
  MMORPGs and emerging business opportunities, challenges and ethical
  implications in metaverses}, Tech. Forcast. Soc. Change 75 (2008) 610--622.
\newblock \href {http://dx.doi.org/10.1016/j.techfore.2007.04.007}
  {\path{doi:10.1016/j.techfore.2007.04.007}}.

\bibitem{Jiang-Ren-Gu-Tan-Zhou-2010-PA}
Z.-Q. Jiang, F.~Ren, G.-F. Gu, Q.-Z. Tan, W.-X. Zhou, {Statistical properties
  of online avatar numbers in a massive multiplayer online role-playing game},
  Physica A 389 (2010) 807--814.
\newblock \href {http://dx.doi.org/10.1016/j.physa.2009.10.028}
  {\path{doi:10.1016/j.physa.2009.10.028}}.

\bibitem{Bogachev-Bunde-2009-EPL}
M.~I. Bogachev, A.~Bunde, {On the occurrence and predictability of overloads in
  telecommunication networks}, EPL (Europhys. Lett.) 86 (2009) 66002.
\newblock \href {http://dx.doi.org/10.1209/0295-5075/86/66002}
  {\path{doi:10.1209/0295-5075/86/66002}}.

\bibitem{Cai-Fu-Zhou-Gu-Zhou-2009-EPL}
S.-M. Cai, Z.-Q. Fu, T.~Zhou, J.~Gu, P.-L. Zhou, {Scaling and memory in
  recurrence intervals of Internet traffic}, EPL (Europhys. Lett.) 87~(4)
  (2009) 68001.
\newblock \href {http://dx.doi.org/10.1209/0295-5075/87/68001}
  {\path{doi:10.1209/0295-5075/87/68001}}.

\bibitem{Jiang-Zhou-Tan-2009-EPL}
Z.-Q. Jiang, W.-X. Zhou, Q.-Z. Tan, {Online-offline activities and game-playing
  behaviors of avatars in a massive multiplayer online role-playing game}, EPL
  (Europhys. Lett.) 88~(4) (2009) 48007.
\newblock \href {http://dx.doi.org/10.1209/0295-5075/88/48007}
  {\path{doi:10.1209/0295-5075/88/48007}}.

\bibitem{Barabasi-2005-Nature}
A.-L. Barab{\'a}si, {The origin of bursts and heavy tails in human dynamics},
  Nature 435 (2005) 207--211.
\newblock \href {http://dx.doi.org/10.1038/nature03459}
  {\path{doi:10.1038/nature03459}}.

\bibitem{Jiang-Xie-Li-Podobnik-Zhou-Stanley-2013-PNAS}
Z.-Q. Jiang, W.-J. Xie, M.-X. Li, B.~Podobnik, W.-X. Zhou, H.~E. Stanley,
  {Calling patterns in human communication dynamics}, Proc. Natl. Acad. Sci.
  U.S.A. 110~(5) (2013) 1600--1605.
\newblock \href {http://dx.doi.org/10.1073/pnas.1220433110}
  {\path{doi:10.1073/pnas.1220433110}}.

\bibitem{Clauset-Shalizi-Newman-2009-SIAMR}
A.~Clauset, C.~R. Shalizi, M.~E.~J. Newman, {Power-law distributions in
  empirical data}, SIAM Rev. 51~(4) (2009) 661--703.
\newblock \href {http://dx.doi.org/10.1137/070710111}
  {\path{doi:10.1137/070710111}}.

\bibitem{Mandelbrot-1983}
B.~B. Mandelbrot, {The Fractal Geometry of Nature}, W. H. Freeman, New York,
  1983.

\bibitem{Eisler-Kertesz-2007-EPL}
Z.~Eisler, J.~Kert{\'e}sz, {Liquidity and the multiscaling properties of the
  volume traded on the stock market}, EPL (Europhys. Lett.) 77~(2) (2007)
  28001.
\newblock \href {http://dx.doi.org/10.1209/0295-5075/77/28001}
  {\path{doi:10.1209/0295-5075/77/28001}}.

\bibitem{Jiang-Guo-Zhou-2007-EPJB}
Z.-Q. Jiang, L.~Guo, W.-X. Zhou, {Endogenous and exogenous dynamics in the
  fluctuations of capital fluxes: An empirical analysis of the Chinese stock
  market}, Eur. Phys. J. B 57~(3) (2007) 347--355.
\newblock \href {http://dx.doi.org/10.1140/epjb/e2007-00174-7}
  {\path{doi:10.1140/epjb/e2007-00174-7}}.

\bibitem{Kolmogorov-1962-JFM}
A.~N. Kolmogorov, {A refinement of previous hypotheses concerning the local
  structure of turbulence in a viscous incompressible fluid at high Reynolds
  number}, J. Fluid Mech. 13~(1) (1962) 82--85.
\newblock \href {http://dx.doi.org/10.1017/S0022112062000518}
  {\path{doi:10.1017/S0022112062000518}}.

\bibitem{VanAtta-Chen-1970-JFM}
C.~W. Van~Atta, W.~Y. Chen, {Structure functions of turbulence in the
  atmospheric boundary layer over the ocean}, J. Fluid Mech. 44~(1) (1970)
  145--159.
\newblock \href {http://dx.doi.org/10.1017/S002211207000174X}
  {\path{doi:10.1017/S002211207000174X}}.

\bibitem{Anselmet-Gagne-Hopfinger-Antonia-1984-JFM}
F.~Anselmet, Y.~Gagne, E.~J. Hopfinger, R.~A. Antonia, {High-order velocity
  structure functions in turbulent shear flows}, J. Fluid Mech. 140 (1984)
  63--89.
\newblock \href {http://dx.doi.org/10.1017/S0022112084000513}
  {\path{doi:10.1017/S0022112084000513}}.

\bibitem{Ghashghaie-Breymann-Peinke-Talkner-Dodge-1996-Nature}
S.~Ghashghaie, W.~Breymann, J.~Peinke, P.~Talkner, Y.~Dodge, {Turbulent
  cascades in foreign exchange markets}, Nature 381~(6585) (1996) 767--770.
\newblock \href {http://dx.doi.org/10.1038/381767a0}
  {\path{doi:10.1038/381767a0}}.

\bibitem{CastroESilva-Moreira-1997-PA}
A.~Castro~e Silva, J.~G. Moreira, {Roughness exponents to calculate
  multi-affine fractal exponents}, Physica A 235~(3) (1997) 327--333.
\newblock \href {http://dx.doi.org/10.1016/S0378-4371(96)00357-3}
  {\path{doi:10.1016/S0378-4371(96)00357-3}}.

\bibitem{Weber-Talkner-2001-JGR}
R.~O. Weber, P.~Talkner, {Spectra and correlations of climate data from days to
  decades}, J. Geophys. Res. 106 (2001) 20131--20144.
\newblock \href {http://dx.doi.org/10.1029/2001GL014170}
  {\path{doi:10.1029/2001GL014170}}.

\bibitem{Kantelhardt-Zschiegner-KoscielnyBunde-Havlin-Bunde-Stanley-2002-PA}
J.~W. Kantelhardt, S.~A. Zschiegner, E.~Koscielny-Bunde, S.~Havlin, A.~Bunde,
  H.~E. Stanley, {Multifractal detrended fluctuation analysis of nonstationary
  time series}, Physica A 316~(1-4) (2002) 87--114.
\newblock \href {http://dx.doi.org/10.1016/S0378-4371(02)01383-3}
  {\path{doi:10.1016/S0378-4371(02)01383-3}}.

\bibitem{Gu-Zhou-2010-PRE}
G.-F. Gu, W.-X. Zhou, {Detrending moving average algorithm for multifractals},
  Phys. Rev. E 82~(1) (2010) 011136.
\newblock \href {http://dx.doi.org/10.1103/PhysRevE.82.011136}
  {\path{doi:10.1103/PhysRevE.82.011136}}.

\bibitem{Grassberger-1983-PLA}
P.~Grassberger, {Generalized dimensions of strange attractors}, Phys. Lett. A
  97~(6) (1983) 227--230.
\newblock \href {http://dx.doi.org/10.1016/0375-9601(83)90753-3}
  {\path{doi:10.1016/0375-9601(83)90753-3}}.

\bibitem{Hentschel-Procaccia-1983-PD}
H.~G.~E. Hentschel, I.~Procaccia, {The infinite number of generalized
  dimensions of fractals and strange attractors}, Physica D 8~(3) (1983)
  435--444.
\newblock \href {http://dx.doi.org/10.1016/0167-2789(83)90235-X}
  {\path{doi:10.1016/0167-2789(83)90235-X}}.

\bibitem{Grassberger-1985-PLA}
P.~Grassberger, {Generalizations of the Hausdorff dimension of fractal
  measure}, Phys. Lett. A 107~(3) (1985) 101--105.
\newblock \href {http://dx.doi.org/10.1016/0375-9601(85)90724-8}
  {\path{doi:10.1016/0375-9601(85)90724-8}}.

\bibitem{Halsey-Jensen-Kadanoff-Procaccia-Shraiman-1986-PRA}
T.~C. Halsey, M.~H. Jensen, L.~P. Kadanoff, I.~Procaccia, B.~I. Shraiman,
  {Fractal measures and their singularities: The characterization of strange
  sets}, Phys. Rev. A 33~(2) (1986) 1141--1151.
\newblock \href {http://dx.doi.org/10.1103/PhysRevA.33.1141}
  {\path{doi:10.1103/PhysRevA.33.1141}}.

\bibitem{Xie-Jiang-Gu-Xiong-Zhou-2015-NJP}
W.-J. Xie, Z.-Q. Jiang, G.-F. Gu, X.~Xiong, W.-X. Zhou, {Joint multifractal
  analysis based on the partition function approach: Analytical analysis,
  numerical simulations and empirical application}, New J. Phys. 17~(10) (2015)
  103020.
\newblock \href {http://dx.doi.org/10.1088/1367-2630/17/10/103020}
  {\path{doi:10.1088/1367-2630/17/10/103020}}.

\bibitem{Chhabra-Sreenivasan-1992-PRL}
A.~B. Chhabra, K.~R. Sreenivasan, {Scale-invariant multiplier distribution in
  turbulence}, Phys. Rev. Lett. 68~(18) (1992) 2762--2765.
\newblock \href {http://dx.doi.org/10.1103/PhysRevLett.68.2762}
  {\path{doi:10.1103/PhysRevLett.68.2762}}.

\bibitem{Jouault-Lipa-Greiner-1999-PRE}
B.~Jouault, P.~Lipa, M.~Greiner, {Multiplier phenomenology in random
  multiplicative cascade processses}, Phys. Rev. E 59 (1999) 2451--2454.
\newblock \href {http://dx.doi.org/10.1103/PhysRevE.59.2451}
  {\path{doi:10.1103/PhysRevE.59.2451}}.

\bibitem{Jiang-Zhou-2007-PA}
Z.-Q. Jiang, W.-X. Zhou, {Scale invariant distribution and multifractality of
  volatility multipliers in stock markets}, Physica A 381 (2007) 343--350.
\newblock \href {http://dx.doi.org/10.1016/j.physa.2007.03.015}
  {\path{doi:10.1016/j.physa.2007.03.015}}.

\bibitem{Muzy-Bacry-Arneodo-1991-PRL}
J.~F. Muzy, E.~Bacry, A.~Arn{\'e}odo, {Wavelets and multifractal formalism for
  singular signals: Application to turbulence data}, Phys. Rev. Lett. 67~(25)
  (1991) 3515--3518.
\newblock \href {http://dx.doi.org/10.1103/PhysRevLett.67.3515}
  {\path{doi:10.1103/PhysRevLett.67.3515}}.

\bibitem{Muzy-Bacry-Arneodo-1993-PRE}
J.~F. Muzy, E.~Bacry, A.~Arn{\'e}odo, {Multifractal formalism for fractal
  signals: The structure-function approach versus the wavelet-transform
  modulus-maxima method}, Phys. Rev. E 47~(2) (1993) 875--884.
\newblock \href {http://dx.doi.org/10.1103/PhysRevE.47.875}
  {\path{doi:10.1103/PhysRevE.47.875}}.

\bibitem{Turiel-Yahia-PerezVicente-2008-JPA}
A.~Turiel, H.~Yahia, C.~J. Perez-Vicente, {Microcanonical multifractal
  formalism - A geometrical approach to multifractal systems: Part I.
  Singularity analysis}, J. Phys. A 41~(1) (2008) 015501.
\newblock \href {http://dx.doi.org/10.1088/1751-8113/41/1/015501}
  {\path{doi:10.1088/1751-8113/41/1/015501}}.

\bibitem{Pont-Turiel-PerezVicente-2009-PA}
O.~Pont, A.~Turiel, C.~J. P{\'e}rez-Vicente, {Empirical evidences of a common
  multifractal signature in economic, biological and physical systems}, Physica
  A 388~(10) (2009) 2025--2035.
\newblock \href {http://dx.doi.org/10.1016/j.physa.2009.01.041}
  {\path{doi:10.1016/j.physa.2009.01.041}}.

\bibitem{Jiang-Xie-Li-Zhou-Sornette-2016-JSM}
Z.-Q. Jiang, W.-J. Xie, M.-X. Li, W.-X. Zhou, D.~Sornette, {Two-state
  Markov-chain Poisson nature of individual cellphone call statistics}, J.
  Stat. Mech.-Theory Exp. 2016~(7) (2016) 073210.
\newblock \href {http://dx.doi.org/10.1088/1742-5468/2016/07/073210}
  {\path{doi:10.1088/1742-5468/2016/07/073210}}.

\bibitem{Shen-Chen-2015-SN}
C.-H. Shen, W.-H. Chen, {Gamers' confidants: Massively Multiplayer Online Game
  participation and core networks in China}, Soc. Networks 40 (2015) 207--214.
\newblock \href {http://dx.doi.org/10.1016/j.socnet.2014.11.001}
  {\path{doi:10.1016/j.socnet.2014.11.001}}.

\bibitem{Malmgren-Stouffer-Campanharo-Amaral-2009-Science}
R.~D. Malmgren, D.~B. Stouffer, A.~S. L.~O. Campanharo, L.~A.~N. Amaral, {On
  universality in human correspondence activity}, Science 325 (2009)
  1696--1700.
\newblock \href {http://dx.doi.org/10.1126/science.1174562}
  {\path{doi:10.1126/science.1174562}}.

\bibitem{Aledavood-Lehmann-Saramaki-2015-FiP}
T.~Aledavood, S.~Lehmann, J.~Saram{\"a}ki, {Digital daily cycles of
  individuals}, Front. in Phys. 3 (2015) 73.
\newblock \href {http://dx.doi.org/10.3389/fphy.2015.00073}
  {\path{doi:10.3389/fphy.2015.00073}}.

\bibitem{Zhou-2012-CSF}
W.-X. Zhou, {Finite-size effect and the components of multifractality in
  financial volatility}, Chaos Solitons Fractals 45~(2) (2012) 147--155.
\newblock \href {http://dx.doi.org/10.1016/j.chaos.2011.11.004}
  {\path{doi:10.1016/j.chaos.2011.11.004}}.

\bibitem{Jiang-Zhou-2008b-PA}
Z.-Q. Jiang, W.-X. Zhou, {Multifractal analysis of Chinese stock volatilities
  based on the partition function approach}, Physica A 387~(19-20) (2008)
  4881--4888.
\newblock \href {http://dx.doi.org/10.1016/j.physa.2008.04.028}
  {\path{doi:10.1016/j.physa.2008.04.028}}.

\bibitem{Zunino-Tabak-Figliola-Perez-Garavaglia-Rosso-2008-PA}
L.~Zunino, B.~M. Tabak, A.~Figliola, D.~G. P{\'e}rez, M.~Garavaglia, O.~A.
  Rosso, {A multifractal approach for stock market inefficiency}, Physica A
  387~(26) (2008) 6558--6566.
\newblock \href {http://dx.doi.org/10.1016/j.physa.2008.08.028}
  {\path{doi:10.1016/j.physa.2008.08.028}}.

\bibitem{deSouza-Queiros-2009-CSF}
J.~de~Souza, S.~M.~D. Queir{\'o}s, {Effective multifractal features of
  high-frequency price fluctuations time series and $\ell$-variability
  diagrams}, Chaos Solitons Fractals 42~(4) (2009) 2512--2521.
\newblock \href {http://dx.doi.org/10.1016/j.chaos.2009.03.198}
  {\path{doi:10.1016/j.chaos.2009.03.198}}.

\bibitem{Zhou-2009-EPL}
W.-X. Zhou, {The components of empirical multifractality in financial returns},
  EPL (Europhys. Lett.) 88~(2) (2009) 28004.
\newblock \href {http://dx.doi.org/10.1209/0295-5075/88/28004}
  {\path{doi:10.1209/0295-5075/88/28004}}.

\bibitem{Theiler-Eubank-Longtin-Galdrikian-Farmer-1992-PD}
J.~Theiler, S.~Eubank, A.~Longtin, B.~Galdrikian, J.~D. Farmer, {Testing for
  nonlinearity in time series: The method of surrogate data}, Physica D 58
  (1992) 77--94.
\newblock \href {http://dx.doi.org/10.1016/0167-2789(92)90102-S}
  {\path{doi:10.1016/0167-2789(92)90102-S}}.

\bibitem{Schreiber-Schmitz-1996-PRL}
T.~Schreiber, A.~Schmitz, {Improved surrogate data for nonlinearity tests},
  Phys. Rev. Lett. 77~(4) (1996) 635--638.
\newblock \href {http://dx.doi.org/10.1103/PhysRevLett.77.635}
  {\path{doi:10.1103/PhysRevLett.77.635}}.

\bibitem{Schreiber-Schmitz-2000-PD}
T.~Schreiber, A.~Schmitz, {Surrogate time series}, Physica D 142~(3-4) (2000)
  346--382.
\newblock \href {http://dx.doi.org/10.1016/S0167-2789(00)00043-9}
  {\path{doi:10.1016/S0167-2789(00)00043-9}}.

\bibitem{Lux-2004-IJMPC}
T.~Lux, {Detecting multifractal properties in asset returns: The failure of the
  ``scaling estimator''}, Int. J. Mod. Phys. C 15~(4) (2004) 481--491.
\newblock \href {http://dx.doi.org/10.1142/S0129183104005887}
  {\path{doi:10.1142/S0129183104005887}}.

\bibitem{Oh-Eom-Havlin-Jung-Wang-Stanley-Kim-2012-EPJB}
G.~Oh, C.~Eom, S.~Havlin, W.-S. Jung, F.~Wang, H.~E. Stanley, S.~Kim, {A
  multifractal analysis of Asian foreign exchange markets}, Eur. Phys. J. B
  85~(6) (2012) 214.
\newblock \href {http://dx.doi.org/10.1140/epjb/e2012-20570-0}
  {\path{doi:10.1140/epjb/e2012-20570-0}}.

\bibitem{Jo-Karsai-Kertesz-Kaski-2012-NJP}
H.-H. Jo, M.~Karsai, J.~Kert{\'{e}}sz, K.~Kaski, {Circadian pattern and
  burstiness in mobile phone communication}, New J. Phys. 14 (2012) 013055.
\newblock \href {http://dx.doi.org/10.1088/1367-2630/14/1/013055}
  {\path{doi:10.1088/1367-2630/14/1/013055}}.

\end{thebibliography}

%

\end{document}